\documentclass[12pt]{article}

\catcode`@=12
\begin{document}
\thispagestyle{empty}
\begin{flushright} 
UCRHEP-T374\\ 
MADPH-04-1377\\
April 2004\
\end{flushright}
\vspace{1.0in}
\begin{center}
{\LARGE	\bf A$_4$ Symmetry and Neutrinos\\
with Very Different Masses\\}
\vspace{1.5in}
{\bf Ernest Ma\\}
\vspace{0.2in}
{\sl Physics Department, University of California, Riverside, 
California 92521 and\\ Phenomenology Institute, Physics Department, 
University of Wisconsin, Madison, Wisconsin 53706}
\vspace{1.5in}
\end{center}
\begin{abstract}\
The $A_4$ model of leptons was proposed originally for nearly degenerate 
neutrino masses.  With minimal modification, it is shown to accommodate as 
well the cases of normal hierarchy and inverted hierarchy. In all cases, 
this new model predicts $\sin^2 2\theta_{atm} = 1$ and $\tan^2 \theta_{sol} = 
0.5$ for $U_{e3} = 0$. It also predicts their deviations from 1 and 0.5 
respectively for $U_{e3} \neq 0$.
\end{abstract}

\newpage
\baselineskip 24pt

Neutrino oscillations are now firmly established \cite{venice}.  The mixing 
matrix $U_{\alpha i}$ linking $(e,\mu,\tau)$ to the neutrino mass 
eigenstates $(\nu_1,\nu_2,\nu_3)$ is also determined to a large extent.  
This information is however not sufficient to fix all the elements of the 
neutrino mass matrix ${\cal M}_\nu$ (assumed here to be Majorana at the 
outset), but is indicative of its possible approximate form in terms of 
a small number of parameters.  For example, if the $Z_2$ symmetry 
$\nu_\mu \leftrightarrow \nu_\tau$ is imposed on ${\cal M}_\nu$, then 
\cite{allp}
\begin{equation}
{\cal M}_\nu = \pmatrix {e & d & d \cr d & a & b \cr d & b & a},
\end{equation}
which is diagonalized by
\begin{equation}
U^T {\cal M}_\nu U = \pmatrix {m_1 & 0 & 0 \cr 0 & m_2 & 0 \cr 0 & 0 & 
m_3},
\end{equation}
where
\begin{equation}
U = \pmatrix {\cos \theta & -\sin \theta & 0 \cr \sin \theta/\sqrt 2 & 
\cos \theta/\sqrt 2 & -1/\sqrt 2 \cr \sin \theta/\sqrt 2 & \cos \theta/\sqrt 2 
& 1/\sqrt 2},
\end{equation}
with
\begin{equation}
\tan 2 \theta = {2 \sqrt 2 d \over e-a-b}.
\end{equation}
In the convention
\begin{equation}
U = \pmatrix {1 & 0 & 0 \cr 0 & c_{23} & -s_{23} \cr 0 & s_{23} & c_{23}} 
\pmatrix {c_{13} & 0 & s_{13} e^{-i\delta} \cr 0 & 1 & 0 \cr -s_{13} 
e^{i\delta} & 0 & c_{13}} \pmatrix {c_{12} & -s_{12} & 0 \cr s_{12} & c_{12} 
& 0 \cr 0 & 0 & 1},
\end{equation}
this means that
\begin{equation}
\theta_{23} = \pi/4, ~~~ \theta_{12} = \theta, ~~~ \theta_{13} = 0, 
\end{equation}
which is of course consistent with present experimental data, i.e.
\begin{equation}
\sin^2 2 \theta_{23} \simeq 1.0, ~~~ \tan^2 \theta_{12} \simeq 0.4, ~~~ 
\sin^2 \theta_{13} < 0.067~(3\sigma).
\end{equation}
Furthermore, it has been shown \cite{allp} that ${\cal M}_\nu$ of Eq.~(1) 
is flexible enough to allow for all possible hierarchies of the neutrino 
mass eigenvalues.  However, since $\theta_{13}$ is required to be zero, 
there can be no $CP$ violation in $U$.

Versatile as it is, the proposed ${\cal M}_\nu$ has a very serious flaw. 
Since each neutrino has its corresponding charged lepton together in a 
weak $SU(2)_L$ doublet, any symmetry imposed on ${\cal M}_\nu$ must affect 
the charged-lepton Dirac mass matrix ${\cal M}_l$ linking $(e,\mu,\tau)$ to 
$(e^c,\mu^c,\tau^c)$.  In particular, the $Z_2$ symmetry $\nu_\mu 
\leftrightarrow \nu_\tau$ implies that there is also a $Z_2$ symmetry 
$\mu \leftrightarrow \tau$ in ${\cal M}_l $.  This requirement is not 
consistent with the implicit assumption of Eq.~(1) that ${\cal M}_l$ is 
diagonal with different values for $m_\mu$ and $m_\tau$.  Attempts to fix 
this problem by choosing other simple symmetries such $Z_3$ \cite{z3} and 
$Z_4$ \cite{z4} necessarily entail additional arbitrary assumptions 
regarding ${\cal M}_l$.  One exception is the recent application \cite{s3} 
of the non-Abelian discrete symmetry $S_3$.  Another is the $A_4$ model 
\cite{mr01,ma02,bmv03,hrsvv}, based on the symmetry of the tetrahedron 
(Plato's fire). This group has three inequivalent one-dimensional 
representations and one three-dimensional representation, allowing 
${\cal M}_l$ to have three independent eigenvalues and in the basis 
where it is diagonal, ${\cal M}_\nu$ has the form
\begin{equation}
{\cal M}_\nu = m_0 \pmatrix {1 & 0 & 0 \cr 0 & 0 & 1 \cr 0 & 1 & 0}.
\end{equation}
This results in three degenerate neutrino masses, but the addition of the 
most general flavor-changing one-loop radiative corrections \cite{bmv03,hrsvv} 
automatically changes it to
\begin{equation}
{\cal M}_\nu = m_0 \pmatrix {1 + \delta_0 + \delta + \delta^* + 2 \delta' &   
\delta'' & \delta''^* \cr \delta'' & \delta & 1 + \delta_0 + (\delta + 
\delta^*)/2 \cr \delta''^* & 1 + \delta_0 + (\delta + \delta^*)/2 & \delta^*}.
\end{equation}
Without loss of generality, $\delta$ may be chosen real by absorbing its 
phase into $\nu_\mu$ and $\nu_\tau$ and $\delta_0$ set equal to zero by 
redefining $m_0$ and the other $\delta$'s.  Remarkably, this ${\cal M}_\nu$ 
has the same form as Eq.~(1), if $\delta''$ is real.  However $\delta''$ is 
in general complex, in which case $\theta_{13}$ becomes nonzero, and the $CP$ 
violation in $U$ is predicted to be maximal \cite{allp,gl03}.  Both results 
are very interesting and sure to be tested in future neutrino experiments.

The neutrino mass matrix of Eq.~(9) predicts three nearly degenerate 
neutrino masses. This is naturally expected from the assumed $A_4$ symmetry. 
Since all the mass splitting comes from radiative corrections, $m_0$ 
should be greater than the square root of the observed $\Delta m^2$ for 
atmospheric neutrino oscillations by at least an order of magnitude.  
This puts a lower bound of about 0.3 eV on $m_0$, which is very relevant 
for neutrinoless double beta decay \cite{beta2} and the large-scale 
structure of the Universe \cite{univ}.  In fact, the experimental upper 
bound on $m_0$ in either case is also about 0.3 eV.

In the following, the $A_4$ model is modified to show that it can also 
accommodate three neutrinos of very different masses, either in the 
normal hierarchy, i.e. $|m_1| < |m_2| < |m_3|$, or the inverted 
hierarchy, i.e. $|m_3| < |m_1| < |m_2|$, with solar neutrino oscillations 
coming from $\nu_1 - \nu_2$ splitting and atmospheric neutrino 
oscillations from the larger splitting of $\nu_3$ with $\nu_{1,2}$. 
It is also shown that the resulting neutrino mixing matrix reduces to that 
of Eq.~(3) in a certain symmetry limit, but with $\tan^2 \theta$ also fixed 
at 1/2. This specific form of $U$ has long been advocated as a desirable 
$ansatz$ \cite{hps} and here is the first example of how it can be derived 
from a complete theory, without arbitrary assumptions regarding its 
charged-lepton sector.  Away from this symmetry limit as parametrized by 
a single (complex) parameter, $U_{e3}$ becomes nonzero, 
$\theta_{23}$ becomes greater or less than $\pi/4$, and $\tan^2 \theta_{12}$ 
moves away from 0.5.  The present experimental bound on $U_{e3}$ means that 
this model has specific predictions for $\theta_{12}$ and $\theta_{23}$, 
which can be tested in future neutrino experiments.

The group theory and representations of $A_4$ are already discussed in 
Ref.~[6].  Briefly, it is the group of even permutations of four objects, 
which is also the symmetry group of the tetrahedron, one of five perfect 
geometric solids known to the ancient Greeks, and associated by Plato 
with the element ``fire''.  It has 12 elements and 4 irreducible 
representations: \underline {1}, \underline {1}$'$, \underline {1}$''$, 
and \underline {3}, with the multiplication rule
\begin{equation}
\underline {3} \times \underline {3} = \underline {1} + \underline {1}' + 
\underline {1}'' + \underline {3} + \underline {3}.
\end{equation}
As such, it is ideal for describing three families of leptons (and quarks).
[Note that $A_4$ is isomorphic to the dihedral group $\Delta (12)$, and is 
a discrete subgroup \cite{disc_sg} of $SO(3)$ as well as $SU(3)$.]  In the 
convention that all fermions are left-handed, the leptons are assumed to 
transform under $A_4$ as follows.
\begin{eqnarray}
&& L_i = (\nu_i,l_i) \sim \underline {3} ~~(i=1,2,3), \\ 
&& l_1^c \sim \underline {1}, ~~~ l_2^c \sim \underline {1}', ~~~ 
l_3^c \sim \underline {1}''.
\end{eqnarray}
Using the Higgs doublets
\begin{equation}
\Phi_i = (\phi^0_i,\phi^-_i) \sim \underline {3} ~~(i=1,2,3),
\end{equation}
the $3 \times 3$ mass matrix linking $l_i$ to $l^c_j$ is given by
\begin{equation}
{\cal M}_l = \pmatrix {h_1 v_1 & h_2 v_1 & h_3 v_1 \cr h_1 v_2 & 
h_2 \omega v_2 & h_3 \omega^2 v_2 \cr h_1 v_3 & h_2 \omega^2 v_3 & 
h_3 \omega v_3},
\end{equation}
where $\omega = e^{2 \pi i/3}$ and $v_i = \langle \phi^0_i \rangle$.  As shown 
in Ref.~[6], for a continuous range of parameter values, the minimum of the 
Higgs potential is given by $v_1=v_2=v_3=v$.  In that case, ${\cal M}_l$ is 
diagonalized by
\begin{equation}
U_L^\dagger {\cal M}_l U_R = \pmatrix {h_1 & 0 & 0 \cr 0 & h_2 & 0 \cr 
0 & 0 & h_3} \sqrt 3 v = \pmatrix {m_e & 0 & 0 \cr 0 & m_\mu & 0 \cr 
0 & 0 & m_\tau},
\end{equation}
where
\begin{equation}
U_L = {1 \over \sqrt 3} \pmatrix {1 & 1 & 1 \cr 1 & \omega & \omega^2 \cr 
1 & \omega^2 & \omega},
\end{equation}
and $U_R$ is the unit matrix.

Instead of using the canonical seesaw mechanism for Majorana neutrino masses 
as in previous versions of the $A_4$ model, the neutrino mass matrix 
${\cal M}_\nu$ is assumed here to come from the naturally small vacuum 
expectation values of heavy Higgs triplets \cite{ms98}.  Let
\begin{equation}
\xi_1 \sim \underline {1}, ~~\xi_2 \sim \underline {1}', 
~~\xi_3 \sim \underline {1}'', ~~\xi_i \sim \underline {3} ~~(i=4,5,6),
\end{equation}
where $\xi_i = (\xi_i^{++},\xi_i^+,\xi_i^0)$. Then ${\cal M}_\nu$ in the 
original basis is given by
\begin{equation}
{\cal M}_\nu = \pmatrix {a+b+c & 0 & 0 \cr 0 & a + \omega b + \omega^2 c & 
d \cr 0 & d & a + \omega^2 b + \omega c},
\end{equation}
where $a$ comes from $\langle \xi_1^0 \rangle$, $b$ from $\langle \xi_2^0 
\rangle$, $c$ from $\langle \xi_3^0 \rangle$, and $d$ from $\langle \xi_4^0 
\rangle$, assuming that $\langle \xi_5^0 \rangle = \langle \xi_6^0 \rangle 
= 0$ which is also a natural minimum of the Higgs potential for a continuous 
range of parameter values.

In the basis where the charged-lepton mass matrix is diagonal, the neutrino 
mass matrix becomes
\begin{equation}
{\cal M}_\nu^{(e,\mu,\tau)} = U_L^\dagger {\cal M}_\nu U_L^* = \pmatrix 
{a + (2d/3) & b - (d/3) & c - (d/3) \cr b- (d/3) & c + (2d/3) & a - (d/3) 
\cr c - (d/3) & a - (d/3) & b + (2d/3)},
\end{equation}
which is the main result of this paper.  For $b=c=d=0$, it reduces to the 
model of Ref.~[8] before radiative corrections. For $b=c$, it reduces to the 
form of Eq.~(1) with the concomitant $\nu_\mu \leftrightarrow \nu_\tau$ 
symmetry and thus the results of Eq.~(6), but with an additional constraint, 
i.e. $e=a+b-d$, or $\tan 2 \theta = -2 \sqrt 2$ in Eq.~(4).  The mixing 
matrix $U$ of Eq.~(3) then becomes
\begin{equation}
U = \pmatrix {\sqrt{2/3} & 1/\sqrt 3 & 0 \cr -1/\sqrt 6 & 1/\sqrt 3 & 
-1/\sqrt 2 \cr -1/\sqrt 6 & 1/\sqrt 3 & 1/\sqrt 2},
\end{equation}
which is a well-known $ansatz$ \cite{hps} with the prediction $\tan^2 
\theta_{12} = 1/2$, but has never been derived from the symmetry of a 
complete theory, without arbitrary assumptions regarding its charged-lepton
sector, until now.

In the basis defined by $U$ of Eq.~(20), i.e.
\begin{eqnarray}
\nu_1 &=& \sqrt {2 \over 3} \nu_e - {1 \over \sqrt 6} (\nu_\mu + \nu_\tau), \\ 
\nu_2 &=& {1 \over \sqrt 3} (\nu_e + \nu_\mu + \nu_\tau), \\ 
\nu_3 &=& {1 \over \sqrt 2} (-\nu_\mu + \nu_\tau),
\end{eqnarray}
the neutrino mass matrix of Eq.~(19) rotates to
\begin{equation}
{\cal M}_\nu^{(1,2,3)} = U^\dagger {\cal M}_\nu^{(e,\mu,\tau)} U^* = 
\pmatrix {m_1 & 0 & m_4 \cr 0 & m_2 & 0 \cr m_4 & 0 & m_3},
\end{equation}
where
\begin{equation}
\pmatrix {m_1 \cr m_2 \cr m_3 \cr m_4} = \pmatrix {1 & -1/2 & -1/2 & 1 \cr 
1 & 1 & 1 & 0 \cr -1 & 1/2 & 1/2 & 1 \cr 0 & -\sqrt 3/2 & \sqrt 3/2 & 0} 
\pmatrix {a \cr b \cr c \cr d}.
\end{equation}
If $m_4 \neq 0$, $\nu_1$ mixes with $\nu_3$, but $\nu_2$ remains the same. 
Let the new mass eigenstates be
\begin{equation}
\nu'_1 = \nu_1 \cos \theta + \nu_3 e^{i\delta} \sin \theta, ~~~ 
\nu'_3 = -\nu_1 e^{-i\delta} \sin \theta + \nu_3 \cos \theta,
\end{equation}
then the new mixing matrix $U$ has elements
\begin{eqnarray}
&& U_{e1} = \sqrt {2 \over 3} \cos \theta, ~~~ U_{e2} = {1 \over \sqrt 3}, ~~~ 
U_{e3} = - \sqrt {2 \over 3} e^{i\delta} \sin \theta, \\
&& U_{\mu 3} = - {1 \over \sqrt 2} \cos \theta + {1 \over \sqrt 6} e^{i\delta} 
\sin \theta = - {1 \over \sqrt 2} \sqrt {1 - {3 \over 2} |U_{e3}|^2} - 
{1 \over 2} U_{e3}.
\end{eqnarray}
Therefore, the experimental constraint \cite{react}
\begin{equation}
|U_{e3}| < 0.16
\end{equation}
implies
\begin{equation}
0.61 < |U_{\mu3}| < 0.77,
\end{equation}
or, using $\sin^2 2 \theta_{atm} = 4 |U_{\mu3}|^2 (1-|U_{\mu3}|^2)$,
\begin{equation}
0.94 < \sin^2 2 \theta_{atm} < 1.
\end{equation}
Similarly, using $\tan^2 \theta_{sol} = |U_{e2}|^2/|U_{e1}|^2$,
\begin{equation}
0.5 < \tan^2 \theta_{sol} < 0.52
\end{equation}
is obtained.  Whereas Eq.~(31) is well satisfied by the current data, 
Eq.~(32) is at the high end of the $90\%~{\rm C.L.}$-allowed range centered 
at $\tan^2 \theta_{sol} \simeq 0.4$ \cite{sol}.

If future experimental measurements persist in getting a value of $\tan^2 
\theta_{sol}$ outside the range predicted by Eq.~(32), one possible 
explanation within the context of this model is that $\langle \xi^0_{5,6} 
\rangle$ are not zero as assumed, but rather $\langle \xi^0_6 \rangle = - 
\langle \xi^0_5 \rangle$, which can be maintained in the Higgs potential 
by postulating the interchange symmetry $\xi_5 \leftrightarrow -\xi_6$ 
in addition to $\phi_2 \leftrightarrow \phi_3$ for the soft trilinear 
$\xi \phi \phi$ terms which break the $A_4$ symmetry.  In that case, the 
neutrino mass matrix of Eq.~(18) has the additional piece
\begin{equation}
\Delta {\cal M}_\nu = \pmatrix {0 & e & -e \cr e & 0 & 0 \cr -e & 0 & 0},
\end{equation}
resulting in
\begin{equation}
\Delta {\cal M}_\nu^{(1,2,3)} = U^\dagger (U_L^\dagger {\cal M}_\nu U_L^*) 
U^* = \pmatrix {0 & 0 & 0 \cr 0 & 0 & m_5 \cr 
0 & m_5 & 0},
\end{equation}
where $m_5 = i \sqrt 2 e$.  Thus Eq.~(24) becomes
\begin{equation}
{\cal M}_\nu^{(1,2,3)} = \pmatrix {m_1 & 0 & m_4 \cr 0 & m_2 & m_5 \cr 
m_4 & m_5 & m_3}.
\end{equation}
If $m_4=0$ but $m_5 \neq 0$, then $\nu_2$ mixes with $\nu_3$ while $\nu_1$ 
remains the same.  The analogs of Eqs.~(27) and (28) are then
\begin{eqnarray}
&& U_{e1} = \sqrt {2 \over 3}, ~~~ U_{e2} = {1 \over \sqrt 3} \cos \theta, ~~~ 
U_{e3} = - {1 \over \sqrt 3} e^{i\delta} \sin \theta, \\
&& U_{\mu 3} = - {1 \over \sqrt 2} \cos \theta - {1 \over \sqrt 3} e^{i\delta} 
\sin \theta = - {1 \over \sqrt 2} \sqrt {1 - 3 |U_{e3}|^2} + U_{e3}.
\end{eqnarray}
Therefore, the experimental constraint \cite{atm}
\begin{equation}
0.91 < \sin^2 2 \theta_{atm} < 1
\end{equation}
implies
\begin{equation}
|U_{e3}| < 0.11,
\end{equation}
and thus
\begin{equation}
0.48 < \tan^2 \theta_{sol} < 0.5.
\end{equation}

If both $m_4$ and $m_5$ are nonzero and all $m_i$'s are real, the eigenvalues 
of Eq.~(35) obey the cubic equation
\begin{equation}
(m_1 - \lambda)(m_2 - \lambda)(m_3 - \lambda) - m_4^2 (m_2 - \lambda) 
- m_5^2 (m_1 - \lambda) = 0.
\end{equation}
For $\lambda_i = m_i + \delta_i$, the eigenstates are approximately given by
\begin{equation}
{\nu_i + \sum_{j \neq i} \epsilon_{ij} \nu_j \over \sqrt {1 + \sum_{j \neq i} 
\epsilon_{ij}^2}},
\end{equation}
where each $\epsilon_{ij}$ is much smaller than 1.  Thus
\begin{eqnarray}
&& U_{e3} = \langle \nu_e|\nu_3 \rangle \simeq {\sqrt 2 \epsilon_{31} + 
\epsilon_{32} \over \sqrt 3}, \\ 
&& U_{\mu3} = \langle \nu_\mu|\nu_3 \rangle \simeq - {\sqrt 3 + \epsilon_{31} 
- \sqrt 2 \epsilon_{32} \over \sqrt 6}, \\ 
&& U_{e1} = \langle \nu_e|\nu_1 \rangle \simeq {\sqrt 2 + \epsilon_{12} \over 
\sqrt 3}.
\end{eqnarray}
Consider first $\nu_3$.  The solutions of Eq.~(41) are
\begin{eqnarray}
&& \delta_3 \simeq {m_4^2 \over m_3-m_1} + {m_5^2 \over m_3-m_2}, \\ 
&& \epsilon_{31} \simeq {m_4 \over m_3-m_1}, ~~~ \epsilon_{32} \simeq 
{m_5 \over m_3-m_2}.
\end{eqnarray}
Consider next $\nu_1$.  The solutions of Eq.~(41) are
\begin{eqnarray}
&& \delta_1 = {1 \over 2} [m_2-m_1-(m_4^2+m_5^2)/(m_3-m_1)] \nonumber \\ 
&& -{1 \over 2} \sqrt {[m_2-m_1-(m_4^2+m_5^2)/(m_3-m_1)]^2 + 4m_4^2 
(m_2-m_1)/(m_3-m_1)}, \\ 
&& \epsilon_{12} = {m_5 \delta_1 \over m_4 (m_1-m_2+\delta_1)}, ~~~ 
\epsilon_{13} = {\delta_1 \over m_4}.
\end{eqnarray}
As a numerical example, consider
\begin{equation}
m_1 = 0.003~{\rm eV}, ~ m_2 = 0.009~{\rm eV}, ~ m_3 = 0.047~{\rm eV}, 
~ m_4 = 0.003~{\rm eV}, ~ m_5 = 0.003~{\rm eV},
\end{equation}
then
\begin{equation}
\epsilon_{31} = 0.068, ~~~ \epsilon_{32} = 0.079, ~~~ \epsilon_{12} = 0.034, 
~~~ \epsilon_{13} = -0.071,
\end{equation}
resulting in
\begin{equation}
U_{e3} = 0.10, ~~~ U_{\mu3} = -0.686, ~~~ U_{e1} = 0.834,
\end{equation}
from which
\begin{equation}
\sin^2 2 \theta_{atm} = 0.996, ~~~ \tan^2 \theta_{sol} = 0.42,
\end{equation}
are obtained.  Accounting also for the mass shifts $\delta_i$, this example 
has
\begin{equation}
\Delta m^2_{atm} = 2.2 \times 10^{-3}~{\rm eV}^2, ~~~ 
\Delta m^2_{sol} = 6.9 \times 10^{-5}~{\rm eV}^2.
\end{equation}
As for $CP$ violation, it is allowed in both Eq.~(24) and Eq.~(35), but is 
not otherwise constrained in this model.

Another possibility of obtaining a value of $\tan^2 \theta_{sol} < 0.5$ in 
Eq.~(24) is through radiative corrections.  Just as Eq.~(8) is radiatively 
corrected to become Eq.~(9), Eq.~(19) may also get corrected so that 
$\nu_1$ mixes with $\nu_2$ in Eq.~(24).  For example, if $b,c,d < a$, then 
combining Eq.~(9) and Eq.~(19) with $b=c$, the analog of Eq.~(4) is given by
\begin{equation}
\tan 2 \theta_{sol} \simeq -2 \sqrt 2 \left[ {b-(d/3)+\delta''a \over 
b-(d/3)-2\delta'a} \right],
\end{equation}
where
\begin{equation}
\delta'' = \delta_{e\mu} + \delta_{e\tau}, ~~~ \delta' = \delta_{ee} - 
{1 \over 2} (\delta_{\mu\mu} + \delta_{\tau\tau}) - \delta_{\mu\tau},
\end{equation}
and $\tan^2 \theta_{sol}\simeq 0.4$ is obtained if $[b-(d/3)+\delta''a]/
[b-(d/3)-2\delta'a] \simeq 0.75$.  Note that this may occur even if 
$\delta_{\alpha \beta}=0$ for $\alpha \neq \beta$, i.e. in the absence of 
flavor changing radiative corrections, 
in contrast to the requirement of Refs.~[8] and [9].

In the Higgs sector, the triplets are very heavy and may all be integrated 
away \cite{ms98}. Their only imprint on the low-energy theory is the neutrino 
mass matrix of Eq.~(19).  On the other hand, the three Higgs doublets are at 
the electroweak mass scale and their Yukawa couplings to the leptons are 
completely determined by the charged-lepton masses as shown in Ref.~[6].  
Phenomenological consequences are discussed in detail there.

In conclusion, a much more predictive version of the $A_4$ model of leptons 
has been proposed.  In the basis where the charged-lepton mass matrix is 
diagonal with three independent eigenvalues (i.e. $m_e$, $m_\mu$, $m_\tau$), 
the neutrino mass matrix is fixed to have the form of Eq.~(19).  Using the 
mixing matrix of Eq.~(20), it becomes Eq.~(24), with 4 parameters 
$m_{1,2,3,4}$.  If $m_4=0$, then $U_{e3}=0$, $\sin^2 2 \theta_{atm} = 1$, 
and $\tan^2 \theta_{sol} = 0.5$, independent of the neutrino mass eigenvalues 
$m_{1,2,3}$.  To obtain the experimentally more favorable result of 
$\tan^2 \theta_{sol} \simeq 0.4$, one possibility is to add $m_5$ according 
to Eq.~(35), another is to include radiative corrections as in Ref.~[8].

I thank Vernon Barger for discussions. 
This work was supported in part by the U.~S.~Department of Energy
under Grant No.~DE-FG03-94ER40837.

\bibliographystyle{unsrt}

\end{document}